\documentclass[12pt]{amsart}
\usepackage{amsmath}
\usepackage{amssymb}
\usepackage{epsfig}
\usepackage{psfrag}
\usepackage{color}
\usepackage{graphics,graphicx}

\begin{document}

\title[]{Combinatorial models of rigidity and renormalization}

\author{J.~Barr\'e} 
\address{Laboratoire J.~A.~Dieudonn\'e\\ 
Universit\'e de Nice-Sophia Antipolis, UMR CNRS 6621 \\ Parc Valrose, 06108 Nice\\ France. }

\subjclass[2000]{}


\thanks{ }
\newtheorem{theorem}{Theorem}[section]
\newtheorem{proposition}[theorem]{Proposition}
\newtheorem{lemma}[theorem]{Lemma}
\newtheorem{corollary}[theorem]{Corollary}
\newtheorem{assumption}[theorem]{Assumption}
\theoremstyle{remark}
\newtheorem{remark}[theorem]{Remark}
\newtheorem{definition}[theorem]{Definition}
\makeatletter
\@addtoreset{equation}{section}
\makeatother
\renewcommand{\theequation}{\arabic{section}.\arabic{equation}}
\newcommand{\cqfd}{{\unskip\kern 6pt\penalty 500
\raise -2pt\hbox{\vrule\vbox to 6pt{\hrule width 6pt
\vfill\hrule}\vrule}\par}}

\begin{abstract}
 We first introduce the percolation problems associated with the graph theoretical concepts of $(k,l)$-sparsity, and make contact with the physical concepts 
 of ordinary and rigidity percolation. We then devise a renormalization transformation for $(k,l)$-percolation problems, and investigate its domain of validity. In particular, we show that it allows an exact solution of $(k,l)$-percolation problems on hierarchical graphs, for
 $k\leq l<2k$. We introduce and solve by renormalization such a model, which has the interesting feature of showing both ordinary percolation and rigidity percolation phase transitions, depending on the values of the parameters.  
\end{abstract}

\maketitle
\section{Introduction}
\label{intro}







Consider an ensemble of bars glued together at joints, around which they can freely rotate, embedded in a two dimensional 
space. The ensemble of bars and joints is naturally associated with a graph: joints are the vertices of the graph, and bars its edges. The question of whether or not this ensemble constitutes a rigid body is a fascinating one. It found a convenient answer with Laman's theorem~\cite{Laman71}, which characterizes rigidity in a purely graph theoretical way \footnote{Throughout this paper, we will be concerned only with \emph{generic rigidity}, which can be reduced to a combinatorial problem. See for instance~\cite{Jacobs98} for a discussion}. It implies for instance that a graph with $N$ vertices must have at least $M=2N-3$ edges to be rigid. Here the number $2$ corresponds to the number of degrees of freedom of a joint and the number $3$ to the number of degrees of freedom of a rigid body, in two dimensions. The concepts of $(k,l)$-sparsity and $(k,l)$-tightness, introduced a long time ago in graph theory \cite{Lorea79}, allow
to generalize the problem~\cite{LeeStreinu07}. For instance, bar-joint rigidity in two dimensions corresponds to $(k,l)=(2,3)$.
Some values of $(k,l)$ have a clear physical meaning in terms of other models of rigidity, others do not, see Sec.~\ref{sec:graph}.

The percolation problem associated with bar-joint rigidity has drawn the attention of physicist's for a long time~\cite{Thorpe83}. It can be stated as follows: take a large lattice, where each site is a joint and each 
bond a bar; consider that each bar is effectively present with probability $p$. 
When $p$ is increased, the system goes from a ``floppy'' phase where only small rigid clusters exist, to a ``rigid'' phase, where there is one macroscopic rigid cluster. In between lies the rigidity percolation transition.
To study this problem, Jacobs, Thorpe, Duxbury and Moukarzel have introduced the so called "pebble game" algorithm, which implements in an efficient way the combinatorial characterization of Laman~\cite{Jacobspebble,Moukarzelpebble}, see also~\cite{Hendrickson92}. The interpretation of the numerical simulations on regular two dimensionnal lattices have fueled some debate, but the following picture seems favored: the rigidity percolation transition in such 2D bar-joints lattices is second order, in a universality class different from ordinary connectivity percolation~\cite{Jacobspebble,Moukarzelpercolation}. We note that rigidity percolation may also be studied on trees or random graphs, in which case it is exactly soluble, and displays a first order transition, at variance with ordinary percolation~\cite{Moukarzelrandomgraphs}. To our knowledge, there exists no analytical theory, even approximate, for the critical exponents of 2D bar-joint rigidity percolation on regular lattices.
  
Except for some special cases, such as ordinary percolation and the bar-joint rigidity percolation described above, percolation problems associated to $(k,l)$ sparsity seem to have received relatively little attention in the physics literature (see Sec.\ref{sec:graph} for a more detailed discussion).
In particular, little is known about the associated percolation transitions and their universality classes. A natural idea to get insight into this problem is renormalization theory. The purpose of this paper is to introduce a renormalization tool to study $(k,l)$-percolation, generalizing the method of \cite{pre09}. In Sec.~\ref{sec:renorm}, we will determine the conditions under which the renormalization procedure is useful. In particular, we will show that it allows a complete solution for some $(k,l)$-rigidity models on hierarchical graphs \cite{kaufman,pre09}. In Sec.\ref{sec:model}, we present and solve explicitly a model of $(3,3)$ percolation (which has a physical interpretation as bodies and bars percolation in two dimensions). We will show that the renormalization transformation on this model has two trivial fixed points, corresponding to the full and empty graph respectively, and two critical fixed points. One fixed point, corresponding to ordinary percolation has three unstable direction, whereas the other one has one unstable direction, and thus governs the critical behavior for almost all values of the parameters. This model provides an example of a system showing both ordinary and rigidity percolation transistions.




\section{Rigidity and graph theory}
\label{sec:graph}

We first briefly recall the graph theoretic description of bar-joints rigidity~\cite{Laman71}.
The basic idea is constraint counting: each joint has two translational degrees of freedom, and each bar, by fixing the distance between two joints, removes one degree of freedom. 
Since a rigid body in two dimensions has three degrees of freedom (two translations and one rotation), an ensemble with $N$ joints needs at least $M=2N-3$ bonds to be rigid. In addition, these bars must be ``well distributed'' among the joints: if there are exactly 
$2N-3$ bonds, there should be no subensemble with $n$ joints containing strictly more than $2n-3$ bonds.

Clearly, the numbers 2 and 3 in this description are adapted to bar-joints rigidity theory in two dimensions, but it is natural to generalize the idea to other pairs of numbers $(k,l)$.
\cite{Tay,LeeStreinu07}. It is also natural to allow for multiple edges in a graph: such graphs are called multigraphs. Although, to our knowledge, they have not appeared in this form in the physics literature, the following definitions about $(k,l)$-rigidity are not new \cite{LeeStreinu07}; we recall them for self-consistency of the article. Throughout the 
article, we shall always consider multigraphs, although we shall keep calling them graphs.

Given a graph $G=(V,E)$, where $V$ represents the set of vertices of $G$ and $E$ the set of its edges, we say that $G'=(V',E')$ is a subgraph of $G$ if $V'\subset V$, $E'\subset E$, and the 
edges in $E'$ connect vertices in $V'$.

\begin{definition} For $k\geq 1$ and $0\leq l\leq 2k-1$, a multigraph $G$ with $N$ vertices and $M$ edges is said to be \\
- $(k,l)$-sparse if every subgraph $G'$ of $G$ with $n>1$ vertices contains at most $m=kn-l$ edges; \\
- $(k,l)$-tight if in addition $M=kN-l$; \\
- $(k,l)$-spanning if there is a $(k,l)$-tight subgraph spanning all vertices of $G$;\\
- $(k,l)$-redundant if it is not $(k,l)$-sparse.\\
Two vertices $A$ and $B$ are said to be $(k,l)$-rigidly connected if there is a $(k,l)$-tight subgraph containing $A$ and $B$. 
\end{definition}

When there is no ambiguity, we shall sometimes omit the $(k,l)$ indices, and talk about sparse, tight, and redundant graphs.

Notice that if $l\geq 2k$, these definitions make no sense: there is no non trivial $(k,l)$-sparse graph in this case, since the simplest graph with $2$ vertices and $1$ edge is not $(k,l)$-sparse. More generally, the edges in a $(k,l)$-sparse multigraph have multiplicity at most $2k-l$.

These graph-theoretic definitions may be immediately translated in the rigidity vocabulary. A $(k,l)$-tight graph is a $(k,l)$ isostatically rigid, or minimally rigid, graph: it has just the right number of constraints to be rigid, and removing any constraint destroys rigidity.  A $(k,l)$-spanning graph is a rigid graph according to the $(k,l)$ rule.

Redundant constraints play an important role in bar and joint rigidity theory. It is straightforward to generalize the concept to $(k,l)$-rigidity. A graph possesses one or more redundant edge
if and only if it is not sparse, ie if it contains a subgraph with $n$ vertices, $m$ edges, and $m>kn-l$. To define and count the number of 
$(k,l)$-redundant edges in a graph $G=(V,E)$, one may proceed as follows:
\begin{enumerate}
\item Start with the empty graph $(V,\emptyset)$, without edges
\item Add the edges of $E$ one at a time 
\item Check if this addition creates a redundant subgraph, with $n$ vertices, $m$ edges, and $m=kn-l+1$. 
\item If this is the case, then reject the newly added edge and add one to the count of redundant constraints
\item Continue until all edges in $E$ have been added. 
\end{enumerate}
This procedure actually follows the steps of the algorithms used to study $(2,3)$-rigidity~\cite{Hendrickson92,Jacobspebble,Moukarzelpebble,JacobsHendrickson}, and more generally  $(k,l)$-rigidity~\cite{LeeStreinu07}. 
A priori, the number of redundant edges computed in this way could depend on the order chosen to add the edges. It is not the case, so that the number of redundant edges
is a well-defined concept. This is also the minimum number of edges that one has to withdraw to make a redundant graph sparse.
In general, there is a freedom of choice in the edges to be removed to turn a redundant graph into a sparse one, but their number is fixed.  



Beyond bar-joint 2D rigidity, which is a physical realization of $(2,3)$-rigidity, $(k,l)$-rigidity encompasses several known situations:
\begin{itemize}
\item $(1,1)$-rigidity corresponds to ordinary percolation, in any dimension.
\item $(3,3)$-rigidity corresponds to the rigidity of systems composed of rigid bodies and bars in 2 dimensions.
\item More generally, $(k,k)$-rigidity is related to rigidity of bodies and bars systems in 
higher dimension \cite{Tay}.
\item $(3,6)$-rigidity is related to the rigidity of bond-bending networks in 3 dimensions \cite{Jacobs98,Jackson07}. In this case $l=6=2k$, which is outside the usual range; 
the definition of sparsity needs to be slightly adapted.
\item $(2,0)$-rigidity corresponds to the graph theoretic concept of 2-orientability, which has been used to study bar-joint rigidity of random networks \cite{Moore}.
\end{itemize}
Clearly, a percolation problem may be associated to each case, by monitoring the largest $(k,l)$-rigid cluster in a graph. Understanding the general properties of these percolation processes, such as the order of the transition, the universality class, etc... is a largely open question, although several cases have been studied in details: among those, ordinary percolation is of course the best known by far.
Percolation for bond-bending networks has been studied, because of its relevance in modeling network glasses and proteins \cite{Thorpeproteins}, but we are not aware of studies on critical exponents. In \cite{Moukarzelpercolation}, C. Moukarzel studies percolation problems on trees and random graphs for various numbers $k$ ($g$ in the notations of 
\cite{Moukarzelpercolation})\footnote{As a tree alone is never rigid, the leafs are usually glued to a busbar to study rigidity for trees.}.
It is also known that when some bars and joints form a rigid component, they behave in every aspect like a rigid body. This idea has actually been numerically exploited \cite{MoukarzelJPA96}. Then one may expect that body-bar ($(k,l)=(3,3)$) and bar-joint 
($(k,l)=(2,3)$) rigidity percolation belong to the same universality class \cite{MoukarzelPRE99}.

Summarizing these informations, one may identify several conjectures and questions:
\begin{enumerate}
\item $(1,1)$ percolation (that is ordinary percolation) and $(2,3)$ percolation on 2D regular lattices belong to two different universality classes: this conjecture seems well documented numerically. 
\item One may also conjecture that $(2,3)$-percolation (bar-joint) and $(3,3)$-percolation (body-bar) on 2D regular lattices belong to
the same universality class.
\item On random graphs, it is known that the threshold for $(2,3)$-percolation is identical to the threshold for $(2,0)$-percolation~\cite{Moore}. One may then wonder if,
on random graphs, the threshold for $(k,l)$-percolation could be independent of $l$, for
$0\leq l \leq 2k-1$.
\end{enumerate}
A more general question is: given for instance a regular lattice in $d$ dimensions; 
what does determine the order of the transition and the universality class of $(k,l)$ percolation? Efficient algorithms, generalizing to $(k,l)$-sparsity the original pebble-game devoted to $(2,3)$ sparsity, have been recently introduced \cite{LeeStreinu07}, so that a numerical investigation of this question is in principle possible. To our knowledge, it has not been undertaken. A natural analytical tool to progress in the studies of these percolation processes is the renormalization group. In section \ref{sec:renorm} we introduce a renormalization transformation,  and we prove that in some sense it is equivalent to study the rigidity properties of the original and renormalized graphs, when the parameter range is restricted to $k\leq l$. This allows to solve exactly models on hierarchical graphs for $k\leq l$; in 
Sec.~\ref{sec:model}, we solve such a model for $k=l=3$.


\section{Renormalization rule for a graph}
\label{sec:renorm}

The renormalization rule we are about to describe is a generalization of the rule introduced on~\cite{pre09} for $(k,l)=(2,3)$. In this case, this rule is intuitive enough so that it does not really require a justification relying on graph theory, see~\cite{pre09} (this rule has been rediscovered independently in~\cite{ThorpeStinchcombe11}). Things are not so obvious for general $(k,l)$, and in particular we have to impose the restriction $k\leq l$. In order to be sure to make correct statements, we feel it is useful to give precise graph theoretical definitions, and to proceed rigorously, not relying only on intuition. 
This certainly makes the article heavier to read, but is necessary to identify for instance the restriction $k\leq l$, which is not completely intuitive.

From now on, we assume $k\leq l<2k$. We will denote $|F|$ the cardinal of an ensemble $F$.

\subsection{Renormalization of a whole graph}
\label{sec:renormgraph}
Consider a multigraph $H=(V,E)$, where $V$ is the set of vertices and $E$ the set of edges. We singularize two sites $A,B\in V$. We would like to replace the graph $H$ by an ``equivalent'' (in a certain sense) graph with only two vertices $A$ and $B$, and a certain number of edges $M_{AB}$ linking $A$ and $B$. We define this renormalization step as follows:\\
Let us call ${\mathcal S}_{AB}$ the ensemble of all $(k,l)$-sparse subgraphs of $H$ containing $A$ and $B$. 
For $h=(V_h,E_h)$ in ${\mathcal S}_{AB}$,  the number of its vertices and edges is respectively $|V_h|$ and $|E_h|$. We define
\[
r_{AB}(H)=\max_{h\in {\mathcal S}_{AB}}[|E_h|-k|V_h|]
\] 
Note that since $h$ is $(k,l)$-sparse, $r_{AB}(H)\leq -l$. Then $M_{AB}$ is given by
\begin{equation}
M_{AB}= \max \left(r_{AB}(H)+2k,0 \right)
\label{eq:renormrule}
\end{equation}
We have $0\leq M_{AB}\leq 2k-l$. The renormalized graph, which contains vertices $A$ and $B$ and $M_{AB}$ edges, is sparse; 
it is tight if and only if $A$ and $B$ are $(k,l)$-rigidly connected in the original graph $H$.
To better understand formula~(\ref{eq:renormrule}), let us explain in details what it means for some particular $(k,l)$.

\begin{figure}
\label{fig:renorm11}
\centering
\includegraphics[width=9cm]{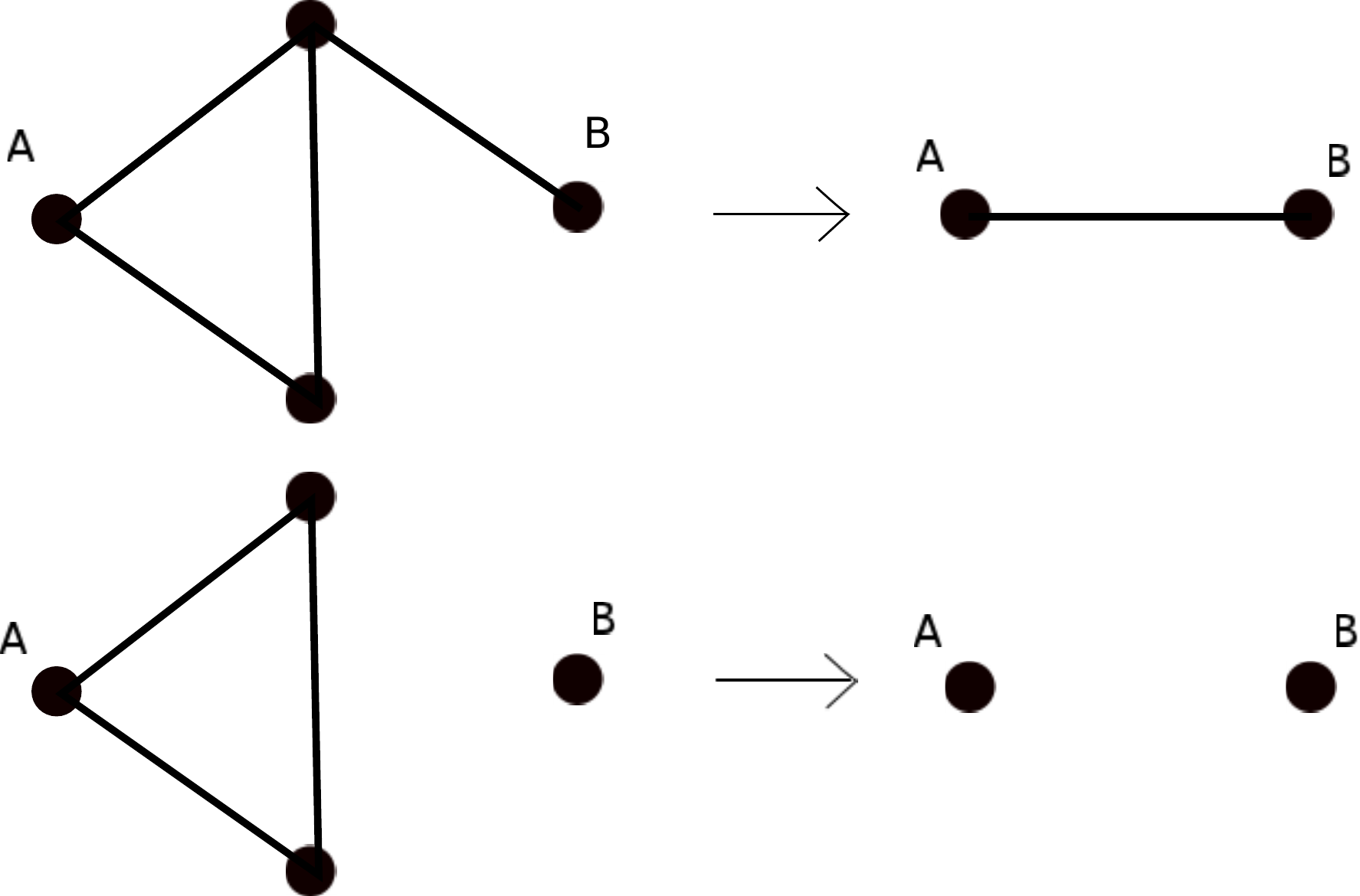}
\caption{Example of renormalization of a simple graph with 4 vertices, $k=l=1$ (ordinary percolation). The upper graph is renormalized into one edge; the lower graph is renormalized into the empty graph. Note that both graphs contain one redundant edge.}
\end{figure}

\begin{figure}
\label{fig:renorm23}
\centering
\includegraphics[width=9cm]{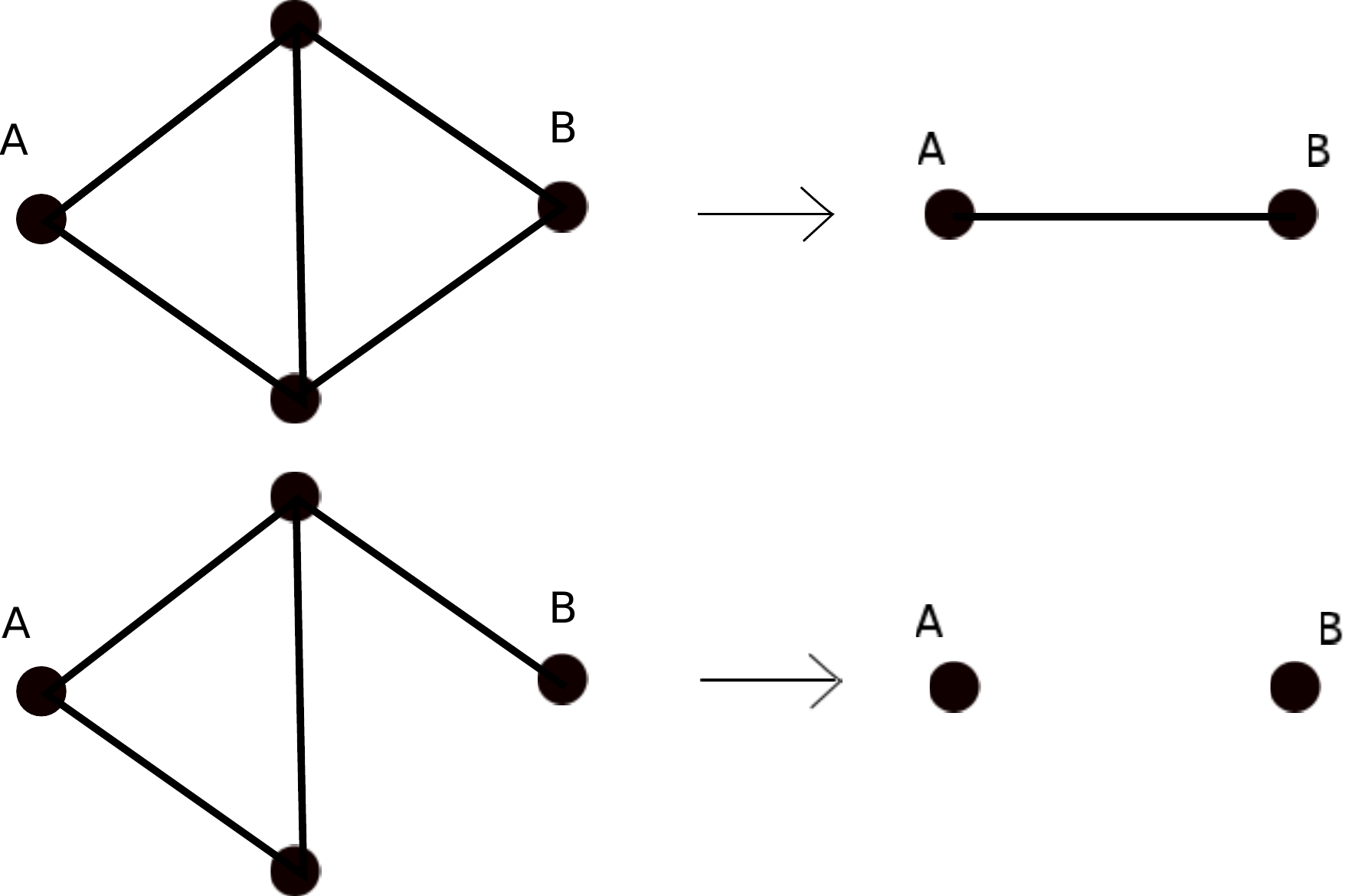}
\caption{Example of renormalization of a simple graph with 4 vertices, $k=2$ and $l=3$ (bar-joint rigidity percolation). The upper graph is renormalized into one edge, as $A$ and $B$ are rigidly connected; the lower graph is renormalized into the empty graph, as $A$ and $B$ are not rigidly connected.}
\end{figure}

\begin{figure}
\label{fig:renorm33}
\centering
\includegraphics[width=9cm]{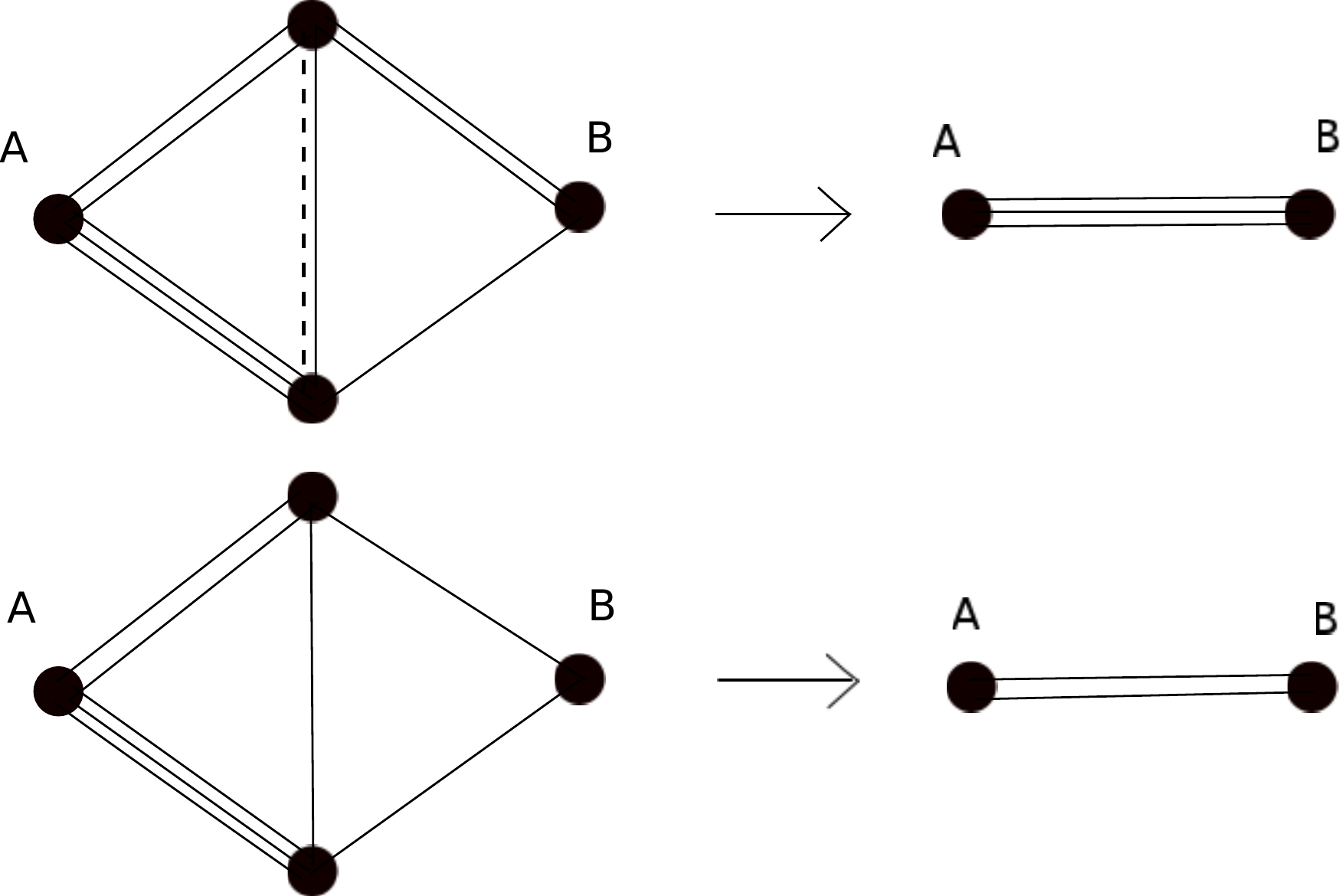}
\caption{Example of renormalization of a simple graph with 4 vertices, $k=3$ and $l=3$ (body-bar rigidity percolation). The upper graph is renormalized into three edges, as the bodies $A$ and $B$ are rigidly connected; the lower graph is renormalized into two edges only.
Note that the upper graph contains one redundant edge, for instance the dashed one.}
\end{figure}

\emph{Ordinary percolation}\\
$(k,l)=(1,1)$ corresponds to the ordinary percolation case. Then $2k-l=1$, and the number of renormalized edges between $A$ and $B$ is $0$ or $1$. Sparse graphs in this case are graphs without any loop.
One sees that a loopless subgraph $h=(V_h,E_h)$ with $|E_h|\geq |V_h|-1$ is necessary for ``percolation'' across the graph. On the elementary ``Wheatstone bridge'' graph of Fig.~\ref{fig:renorm11}, it is easy to see that this coincides with the usual renormalization prescription: replace the graph by an edge if and only if $A$ and $B$ are connected.

\emph{Bar-joint rigidity percolation}\\
In this case $(k,l)=(2,3)$. Again $2k-l=1$, and the number of renormalized edges between $A$ and $B$ is $0$ or $1$. The rule says in this case: ``Renormalize the graph by a edge if and only if $A$ and $B$ are rigidly connected (ie belong to the same rigid cluster)''.
This is intuitive, and this is the rule which has been used in \cite{pre09}.
Examples are given on Fig.~~\ref{fig:renorm23}.

\emph{Body-bar rigidity percolation}\\
In this case $(k,l)=(3,3)$. $2k-l=3$, so that the number of renormalized edges between $A$ and $B$ is $0,1,2$ or $3$.
$A$ and $B$ have to be seen as rigid bodies, with three degrees of freedom each, and three edges are necessary for a rigid connection between them. Examples are given on Fig.~~\ref{fig:renorm33}.


The following property is useful to effectively compute renormalized graphs.
\begin{proposition}\label{prop1}
Let $H=(V,E)$ be sparse, with $A,B \in V$. Let $e$ be an edge such that $H'=(V,E\cup \{ e \})$ is not sparse. 
Then $r_{AB}(H)=r_{AB}(H')$. 
\end{proposition}
As a consequence of this proposition, the renormalizations of $H$ and $H'$ contain the same number of edges. In other words, adding redundant edges to a graph $H$ does not modify its renormalization. We will use this as a tool to compute renormalized graphs in 
Sec.~\ref{sec:model}.

{\bf Proof:} 
Clearly, $r_{AB}(H)\leq r_{AB}(H')$, since an element of ${\mathcal S}_{AB}(H)$ is also in a natural way an element of ${\mathcal S}_{AB}(H')$.

We call $u$ and $v$ the two vertices connected by the edge $e$; $u,v \in V$.

Let us now consider $g'=(V',E')$ an element of  ${\mathcal S}_{AB}(H')$, such that the maximum is attained in the definition of 
$r_{AB}(H')$.
Thus $|E'|-k|V'|=r_{AB}(H')$. We want to construct a subgraph $g''=(V'',E'')$ in ${\mathcal S}_{AB}(H)$ such that
 $|E''|-k|V''|\geq r_{AB}(H')$. This will prove $r_{AB}(H)\geq r_{AB}(H')$.
 We may assume that $e\in E'$, otherwise it is enough to take $g''=g'$. Thus the vertices $u$ and $v$ are in $V'$.
 
 Let $G'_0=(V_0,E_0\cup \{e\})$ be a minimal redundant subgraph of $H'$. $V_0$ contains $u$ and $v$ the vertices connected by $e$, and $G_0=(V_0,E_0)$ is a tight subgraph of $H$. 
We define $V''=V' \cup V_0$ and $g''=(V'',E'')$ as the subgraph of $H$ induced by the vertices $V' \cup V_0$. $g''$ is an element of  ${\mathcal S}_{AB}(H)$. $g''$ does not contain $e$, but we will show that $|E''|-k|V''|\geq r(H')$.

Assume first $V_0\subset V'$. In this case $V''=V'$. Then there is one edge $e_0$ in $E_0\cup \{e\}$ that is not in $E'$, otherwise $g'$ would not be sparse. 
$e_0$ is different from $e$, thus $e_0$ is also in $E''$. We conclude $|E''|-k|V''|\geq |E'|-k|V'|=r_{AB}(H')$.

We assume now that $V_0\backslash V'$ is not empty. We have the relations:
\[
V''=V'\sqcup (V_0\backslash V')~,~E'' \supset (E'\backslash \{e\}) \sqcup (E_0\backslash E')
\]
where $\sqcup$ is a disjoint union. Indeed, $e\in E'$ and $e\notin E''$. Thus
\begin{equation}
|V''|=|V'|+|V_0\backslash V')|~,~|E''|\geq |E'|-1+|E_0\backslash E'|
\label{eq1}
\end{equation}
Let $G_1=(V_0\cap V', E_0\cap E')$ the subgraph defined by the intersection of $V'$ and $V_0$. $G_1$ is a subgraph of $H$, so it is sparse; it is also a proper subgraph of $G_0$, and it contains vertices $u$ and $v$, so it is not tight, otherwise it would contradict the minimality of $G'_0$: indeed, $(V_0\cap V', E_0\cap E'\cup \{e\})$ would be a redundant subgraph of $H'$, smaller than $G'_0$.
We conclude that
\begin{equation}
\label{eq:interm}
|E_0\cap E'| < k|V_0\cap V'| - l
\end{equation}
We use now $V_0=(V_0\cap V')\sqcup (V_0\backslash V')$  and 
$E_0=(E_0\cap E')\sqcup (E_0\backslash E')$.
Since $G_0$ is tight, we have
\begin{equation}
|E_0\cap E'|+|E_0\backslash E'|-k|V_0\cap V'|-k|V_0\backslash V'|-l=0
\end{equation}
Using (\ref{eq:interm}), we get
\[
|E_0\backslash E'|-k|V_0\backslash V'|>0
\]
Putting this together with (\ref{eq1}), we have finally
\[
|E''|-k|V''|\geq |E'|-k|V'| -1 + |E_0\backslash E'|-k|V_0\backslash V'|\geq |E'|-k|V'| =r_{AB}(H')
\]
\cqfd 

\subsection{Renormalization of a subgraph}
\label{sec:renormsubgraph}
\begin{figure}
\label{fig:renormalizability}
\centering
\includegraphics[width=9cm]{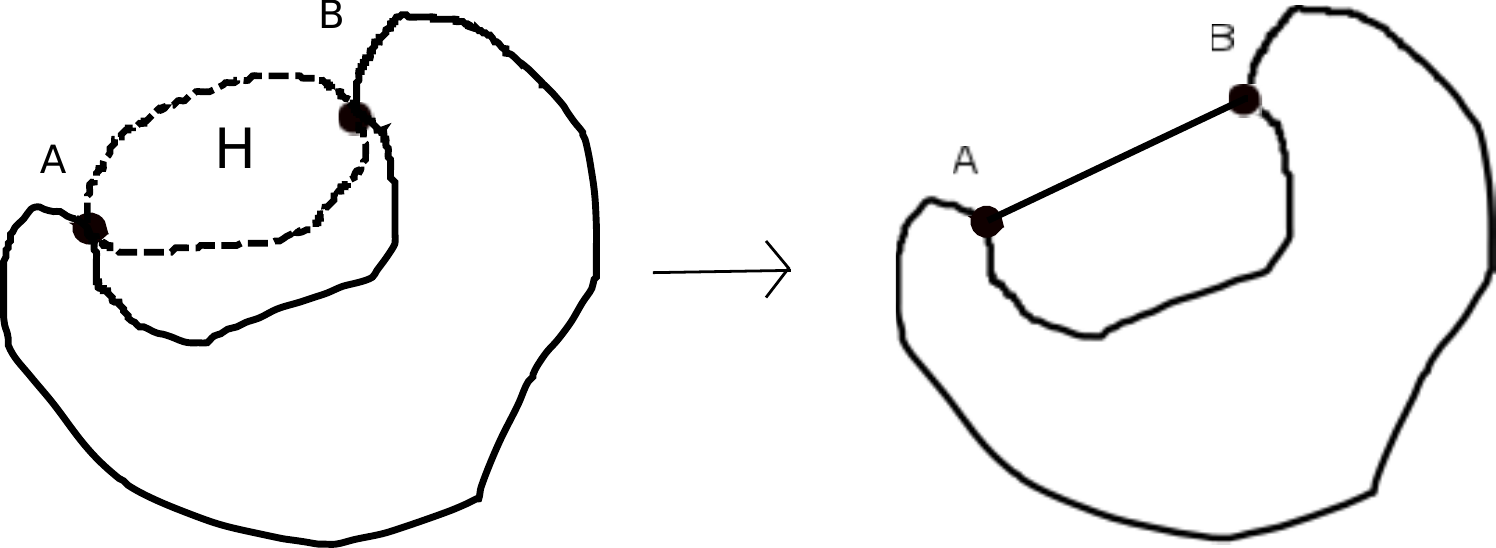}
\caption{On the left is schematized a graph $G$; the subgraph $H$ satisfies the renormalizability condition. On the right, $H$ has been renormalized into one edge, the rest of the graph is not modified.}
\end{figure}

We have seen how to replace a graph by a certain number of renormalized edges.
Now, we will see why and when it is licit to perform this renormalization step on a subgraph of a bigger graph. Consider a graph 
$G=(V,E)$, $V_H\subset V$ and $H=(V_H,E_H)$ the subgraph of $G$ induced by the vertices $V_H$. 
We will say that $H$ satisfies the \emph{renormalizability condition} if there exist two vertices $A$ and $B$ such that 
all edges in $E$ linking $H$ to the rest of the big graph $G$ are connected to $A$ or $B$; see Fig.\ref{fig:renormalizability}. In the following, we always assume this condition is satisfied.

For any subgraph of $g\subset G$, we can define the renormalized graph $R(g)$, as follows: the part of $g$ which is a subgraph of $H$ is renormalized according to the rule detailed in \ref{sec:renormgraph}; the remaining part of $g$ is left unchanged. Note that if $A$ or $B$, or both, do not belong to $g$, then the renormalized part of $g$ contains no edge.

 Let us call $\tilde{G}=R(G)=(\tilde{V},\tilde{E})$ the graph $G$ where the subgraph $H$ has been renormalized. 
We would like to show that it is in some sense equivalent to study $(k,l)$-percolation on $G$ and $\tilde{G}$. For this purpose, we will prove the following:

\begin{proposition}\label{prop2}
In the setting described above:\\
i) $A$ and $B$ are $(k,l)$-rigidly connected in $G$ if and only if they are $(k,l)$-rigidly connected in $\tilde{G}$.\\
ii) The number of redundant edges in $G$ and $\tilde{G}$ are related through the formula
\[
n_{red}(G) = n_{red}(\tilde{G})+ n_{red}(H) 
\]
where we have written $n_{red}(F)$ for the number of redundant edges in a graph $F$.
\end{proposition}
This last formula allows then to replace the problem of computing the number of redundant
edges in $G$ by the the problem of computing the number of redundant
edges in the renormalized graph $R(G)=\tilde{G}$.

 
 Before proving \ref{prop2}~i) and ii), we state and prove a few lemmas.
 
\begin{lemma} \label{lemma1}
Let $G=(V,E)$ and $R(G)=(\tilde{V},\tilde{E})$ be as above. Let $\tilde{g}=(\tilde{v},\tilde{e})$ be a subgraph of $R(G)$, and let us call $\tilde{h}$ its renormalized part; $\tilde{h}$ contains $\tilde{m}$ edges. Then there exists $g$ subgraph of $G$, such that\\ 
i) $\tilde{g}=R(g)$\\
ii) If $A$ or $B$ do not belong to $\tilde{v}$, then, calling $h=(V_h,E_h)$ the intersection of $g$ with $H$, $E_h$ is empty (ie $h$ has no edge).\\
iii) If $A$ and $B$ belong to $\tilde{v}$, then $h=(V_h,E_h)$ is sparse and satisfies
\[
\tilde{m}=|E_h|-k|V_h|+2k
\]
\end{lemma}
{\bf Proof}: Note first that $\tilde{h}$ has at most two vertices, $A$ and $B$. If $A$ or $B$ do not belong to $\tilde{g}$, or if $\tilde{m}=0$ then we choose for the vertices of $h$ the same vertices as $\tilde{h}$, and put no edge. 

If $\tilde{m}>0$, we take $h=(V_h,E_h)$ an element of ${\mathcal S}_{AB}(H)$ such that $|E_h|-k|V_h|+2k=\tilde{m}$, and
$h$ minimal in the sense that no proper subgraph of $h$ has the same properties. Then the renormalization of
$h$ produces exactly $\tilde{m}$ edges between $A$ and $B$. Indeed, suppose it produces $m'>\tilde{m}$ such edges;
then there exists $h'=(V'_h,E'_h)$ subgraph of $h$, in ${\mathcal S}_{AB}(H)$, such that $|E'_h|-k|V'_h|+2k=m'$. By removing edges,  we would construct a proper subgraph of $h$ with the same properties as $h$.

To build $g$, we complete $h$ in both cases by the part of $\tilde{g}$ not concerned by the renormalization.
\cqfd

\begin{lemma}\label{lemma2} 
Let $g=(V_g,E_g)$ be a $(k,l)$-redundant subgraph of $G$, minimal (in the sense that no proper subgraph of $g$ is $(k,l)$-redundant). Then i) or ii) is true:\\
i) $g$ is a subset of $H$ (that is $g$ is included in the part of $G$ to be renormalized).\\
ii) $\tilde{g}=R(g)$ is a $(k,l)$-redundant subgraph of $\tilde{G}=R(G)$ (that is the image of $g$ by renormalization is still 
$(k,l)$-redundant).
\end{lemma}
{\bf Proof}: Suppose i) is not true, and $A,B\in g$. Let us call $h=(V_h,E_h)$ the intersection of $g$ and $H$, and $\tilde{h}=R(h)$. 
We call $\tilde{m}$ the number of edges of $\tilde{h}$. The set of vertices and edges of $\tilde{g}$ are called $V_{\tilde{g}}$ and 
$E_{\tilde{g}}$ respectively.
Then $|E_g|=|E_{\tilde{g}}|-\tilde{m}+|E_h|$ and $|V_g|=|V_{\tilde{g}}|+|V_h|-2$. Thus
\begin{equation}
|E_g|-k|V_g|+l = |E_{\tilde{g}}| -k |V_{\tilde{g}}| +l + |E_h|-k|V_h|+2k
\label{eq:l21}
\end{equation}
Now, $h$ is (k,l)-sparse since $g$ is a minimal redundant graph. Thus $h\in {\mathcal S}_{AB}(h)$, and  
\begin{equation}
\tilde{m}\geq     |E_h|-k|V_h|+2k   
\label{eq:l22}
\end{equation}
Putting together Eqs.~(\ref{eq:l21}) and (\ref{eq:l22}), an using $|E_g|-k|V_g|+l>0$ we get
\begin{equation}
|E_{\tilde{g}}| -k |V_{\tilde{g}}| +l >0
\end{equation}
This proves that ii) is true.

If  i) is not true and neither $A$ nor $B$ are in $g$, then $h$ is actually empty: otherwise, $g$ would be a minimal redundant subgraph
with two disconnected components, which is impossible. Then the renormalization does not modify $g$ and ii) is true.

If  i) is not true and $A$ or $B$, but not both, are in $g$, then either $h$ has no edge, and we are done since $g$ is not modified by renormalization, or $g$ is a minimal redundant graph made of two components that share only one vertex. It is easy to show by enumerating vertices and edges that this is impossible as soon as $l\geq k$. This ends the proof.
\cqfd

\begin{figure}
\centering
\includegraphics[width=9cm]{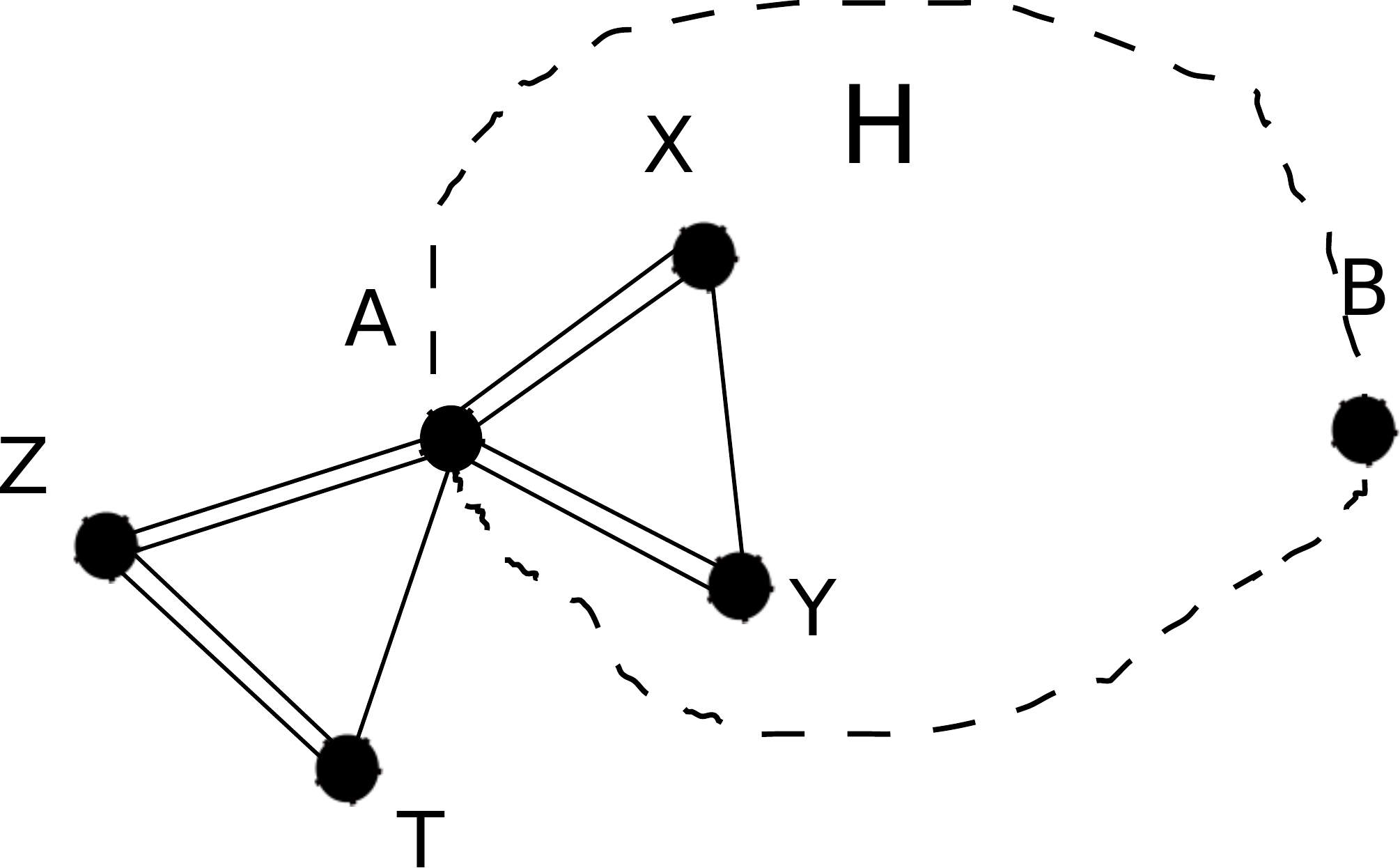}
\caption{For $(k,l)=(2,1)$, the subgraph $H$ (induced by vertices $(A,B,X,Y)$) does not contain any redundant edge, and is renormalized into a subgraph with a 
single edge between $A$ and $B$. The renormalized subgraph is sparse. However, the subgraph induced by the vertices 
$(A,X,Y,Z,T)$ is redundant. \label{fig:contrex}}
\end{figure}

{\bf NB}: Notice that for $0\leq l<k$, this lemma is not true. Fig.~\ref{fig:contrex} shows a counter example for $(k,l)=(2,1)$. It is not 
clear to us how it is possible to define a useful renormalization transformation in this case.

\begin{lemma} \label{lemma3}
Let $\tilde{g}=(\tilde{v},\tilde{e})$ be a $(k,l)$-redundant subgraph of $\tilde{G}$. Then there exists $g$, $(k,l)$-redundant subgraph of $G$ such that $\tilde{g}=R(g)$ (that is: $\tilde{g}$ is the image by renormalization of a $(k,l)$-redundant subgraph of $G$).
\end{lemma}
{\bf Proof}: 
If $\tilde{e}$ does not contain any $A-B$ edge, then it is enough to take $g=(\tilde{v},\tilde{e})$.

We assume now that $\tilde{e}$ contain $\tilde{m}>0$ edges linking $A$ and $B$. This implies that $M_{AB}(H)>0$ (since $M_{AB}(H)$ is the number of $A-B$ edges in $\tilde{G}$).

Using the same reasoning as in Lemma~\ref{lemma1}, we construct $h=(V_h,E_h)$ an element of ${\mathcal S}_{AB}(H)$ such that 
$|E_h|-k|V_h|+2k=\tilde{m}$ and such that
the renormalization of $h$ produces exactly $\tilde{m}$ edges between $A$ and $B$.

Define now $g=(V_g,E_g)$, with $V_g=\tilde{v}\cup V_h$ and $E_g$ contains $E_h$ and 
$\tilde{e}$ minus the $\tilde{m}$ edges between $A$ and $B$. Then $R(g)=\tilde{g}$, and
\[
|E_g|-k|V_g|+l=|\tilde{e}|-\tilde{m}+|E_h|-k|\tilde{v}|+2k-k|V_h|+l=|\tilde{e}|-k|\tilde{v}|+l>0
\]
Then $g$ is redundant.\cqfd

\medskip

{\bf Proof of} \ref{prop2} i):\\
\emph{Rigid connection in $G$} $\Rightarrow$ \emph{rigid connection in $\tilde{G}$}: Suppose $A$ and $B$ are $(k,l)$-rigidly connected in $G$. Then there exists $G_1=(V_1,E_1)$ a $(k,l)$-tight subgraph of $G$ containing $A$ and $B$. Let us call $H_1=(V_{H_1},E_{H_1})$ the intersection of $G_1$ and the subgraph to be renormalized $H$; 
$\tilde{G_1}=R(G_1)=(\tilde{V}_1,\tilde{E}_1),\tilde{H_1}=R(H_1)=(\tilde{V}_{H_1},\tilde{E}_{H_1})$ the subgraphs of $\tilde{G}$ 
images of $G_1,H_1$ by renormalization; $\tilde{H_1}$ has only two vertices $A$ and $B$, and a certain number of edges between them. 

We have 
\[
|V_1|=|\tilde{V}_1|+|V_{H_1}|-2~;~|E_1|=|\tilde{E}_1|+|E_{H_1}|-|\tilde{E}_{H_1}|
\]
$H_1$ is a subgraph of $G_1$ which is tight, so that $H_1$ is sparse. The subgraph of $G_1$ defined by the vertices
$V_1\backslash V_{H_1} \cup \{A,B\}$ is also sparse. This yields the two inequalities  
\[
k|V_{H_1}|-2k \leq |E_{H_1}| \leq k |V_{H_1}|-l
\]
Then, by renormalization of $H_1$, we have
\[
|\tilde{E}_{H_1}|=|E_{H_1}|-k|V_{H_1}|+2k
\]
From this and the fact that $G_1$ is tight, we obtain finally $|\tilde{V}_1|=k|\tilde{E}_1|-l$. 

Now we want to prove that $\tilde{G}_1$ is sparse. Let $\tilde{g}=(\tilde{v},\tilde{e})$ be a subgraph of $\tilde{G}_1$. We have to show that $|\tilde{v}|-k|\tilde{e}|+l\leq 0$.
By Lemma~\ref{lemma1}, we construct $g=(v,e)$. $h=(V_h,E_h)$ is the intersection of $g$ with $H_1$, and $\tilde{m}$ is the number of edges in the renormalization of $h$. Then using point \emph{iii)} of lemma~\ref{lemma1} and
\[
|v|=|\tilde{v}|+|V_h|-2~;~|e|=|\tilde{e}|-\tilde{m}+|E_h|~,
\]
we conclude
\[
|\tilde{e}|-k |\tilde{v}| +l \leq 0
\]
This proves that $\tilde{G}_1$ is $(k,l)$-tight, so that $A$ and $B$ are rigidly connected in $\tilde{G}$.\\

\emph{Rigid connection in $\tilde{G}$} $\Rightarrow$ \emph{rigid connection in $G$}: We assume now that $A$ and $B$ are rigidly connected in $\tilde{G}$, and want to show this is true also in $G$.
Let $\tilde{G}_1=(\tilde{V}_1,\tilde{E}_1)$ be a $(k,l)$-tight subgraph of $\tilde{G}$ containing $A$ and $B$. Using Lemma~\ref{lemma1}, we construct $G_1=(V_1,E_1)$ subgraph of $G$, and call $H_1=(V_{H_1},E_{H_1})$ the intersection of $G_1$ with $H$. 
First, reasoning as above, we have  $|E_1|=k|V_1|-l$.\\
We now have to prove that $G_1$ does not contain any $(k,l)$-redundant subgraph. Suppose it is not the case, and call $g_1$ a minimal redundant subgraph of $G_1$. Then we have the alternative of Lemma~\ref{lemma2}. i) cannot be true because $g_1$ would be a subgraph of $H_1$, which is sparse by Lemma~\ref{lemma1}. ii) cannot be true either, because we would have constructed a redundant subgraph of 
$\tilde{G}_1$, which is tight by hypothesis. We conclude that $G_1$ is tight, and that $A$ and $B$ are rigidly connected in $G_1$.\\

{\bf Proof of} \ref{prop2} ii):\\
We want to prove the formula
\begin{equation}
n_{red}(G)=n_{red}(\tilde{G})+n_{red}(H)
\label{eq:red}
\end{equation}
To prove this, we follow the algorithm defining $n_{red}(G)$, as described in Sec.~\ref{sec:graph}. Starting from the vertices $V$ and no edge, we add the edges in $E$ one by one, starting with the edges which are in $H$, the subgraph to be renormalized. If the newly added edge is redundant, we discard it and add one to the count of 
$n_{red}(G)$. This way, we construct a sequence of sparse graphs 
$$
G_0=(V,\emptyset),G_1,\ldots,G_{M_H},\ldots,G_{M}=(V,E_{M})~,
$$ 
where $M_H=|E_H|$ and $M=|E|$. We have $n_{red}(G)=M-|E_{M}|$. $E_M$ contains the edges in $E$ which have not been found redundant in the sequential edge addition process. 

Consider the associated sequence of renormalized graphs 
$$
\tilde{G}_0=R(G_0),\tilde{G}_1,\ldots,\tilde{G}_{M_H},\ldots,\tilde{G}_M=R(G_M)
$$ 
This is a sequence of sparse graphs thanks to lemma~\ref{lemma3}, and along this sequence, all edges in $\tilde{E}$ are added one by one. We will show we can use this sequence to count the number of redundant edges in $\tilde{G}$.\\

{\bf Adding the edges of $H$}:\\
When adding one by one the edges of $H$, exactly $n_{red}(H)$ edges are discarded, and are added to the count of $n_{red}(G)$. 
After addition of all edges of $H$, $\tilde{G}_{M_H}$ contains exactly $M_{AB}(H)$ edges between vertices $A$ and $B$. 
Along this sequence of edges additions, $(\tilde{G}_n)_{n\leq M_H}$ contains only a number smaller than $2k-l$ of edges connecting $A$ and $B$, and no other edge. Thus, the count of $n_{red}(\tilde{G})$ remains~$0$.\\

{\bf Adding an edge $e\in E\backslash E_H$ (ie an edge in $G$, but not in $H$) }:\\
We start with the graphs $G_n=(V,E_n)$ and $\tilde{G}_n=(\tilde{V},\tilde{E}_n)$, with $n\geq M_H$.
In this case, trying to add an edge in $G$ corresponds to trying to add the same edge in $\tilde{G}$. We will show that
$e$ makes either both $G_n$ and $\tilde{G}_n$ redundant, or none of the two. \\

\emph{Case 1}: $(V,E_n\cup \{ e \})$ is not redundant: the added edge is accepted, $G_{n+1}=(V,E_n\cup \{ e \})$, and the count of
$n_{red}(G)$ is not modified.\\
 Then $(\tilde{V}, \tilde{E}_n\cup \{ e \})$ is not redundant: otherwise, by lemma~\ref{lemma3}, it would be the image by renormalisation of a redundant subgraph of $(V,E_n\cup \{ e \})$, which is impossible. Then $\tilde{G}_{n+1}=(\tilde{V},\tilde{E}_n\cup \{ e \}))$, and
 neither the count of $n_{red}(G)$ nor the count of $n_{red}(\tilde{G})$ are modified.\\

\emph{Case 2}: $(V,E_n\cup \{ e \})$ is redundant: the added edge is discarded, $G_{n+1}=(V,E_n)=G_n$, and the count of $n_{red}(G)$ is increased by~$1$.\\
Then $(\tilde{V}, \tilde{E}_n\cup \{ e \})$ is redundant. Indeed, let $g$ by a minimal redundant subgraph of $(V,E_n\cup \{ e \})$.
It cannot be included in $H$, as it contains the edge $e$. Then, by lemma~\ref{lemma2}, $R(g)$ is a redundant subgraph of
 $(\tilde{V}, \tilde{E}_n\cup \{ e \})$. Thus the count of $n_{red}(\tilde{G})$ is also increased by~$1$, and $\tilde{G}_{n+1}=\tilde{G}_n$.\\

Repeating this until all edges in $E\backslash E_H$ have been added proves formula~(\ref{eq:red}).\cqfd

Notice that formula~(\ref{eq:red}) is wrong for the graph of Fig.~\ref{fig:contrex}, with $(k,l)=(2,1)$. 

\subsection{Hierarchical graphs}

\begin{figure}
\centering
\includegraphics[width=9cm]{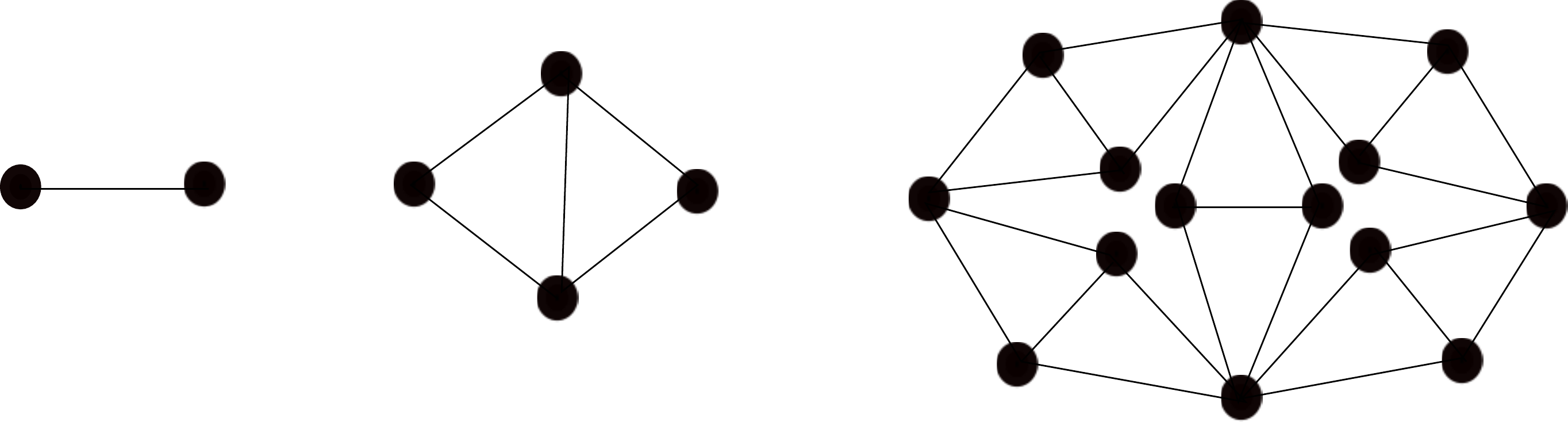}
\caption{A hierarchical graph constructed from a "Wheatstone bridge" elementary cell. From left to right, the level is $t=0, ~t=1, ~t=2$. 
This is the graph which is used in Section~\ref{sec:model}, where in addition each edge may have multiplicity up to $3$.
\label{fig:hierarchical}}
\end{figure}

If it is possible to iterate the renormalization transformation we have just defined until the graph becomes trivial, then the problems of
percolation and counting the number of redundant edges are exactly solved. It is indeed possible to define some graphs on which this procedure can be carried out completely, thus providing exactly solvable models of $(k,l)$ percolation that go beyond trees and random graphs. These graphs are called ``hierarchical graphs'' 
\cite{Berker79}, and are defined as follows. We start
from two vertices, connected by one edge. The graph is then constructed
iteratively; at each step, all edges are replaced by a given
elementary cell. From each type of elementary cell, one thus constructs a hierarchical graph.
A graph where this replacement procedure has been iterated $t$ times will be called a level $t$ hierarchical graph.
An example is given on Fig.~\ref{fig:hierarchical}. 

Consider now within a hierarchical graph one elementary cell between vertices $A$ and $B$. It clearly satisfies the renormalizability condition of Sec.\ref{sec:renormsubgraph}: all edges going outside of the cell 
are connected to the outer vertices $A$ and $B$, whereas the inner vertices are only linked within the cell.
Thus, the renormalization procedure described in Sec.~\ref{sec:renormsubgraph} is exact 
when applied to an elementary cell. Furthermore, if the renormalization procedure is 
applied to all elementary cells of a level $t$ hierarchical graph, the resulting renormalized graph is again a hierarchical graph, of level $t-1$. These remarks explain why the renormalization procedure allows to solve exactly $(k,l)$-rigidity models on hierarchical graphs.

\section{An exactly solved model}
\label{sec:model}

In this section, we apply the formalism developed in the previous section to a body-bar rigidity model: that is we study $(3,3)$ percolation
on a multigraph $G$, with $N$ vertices and $M$ edges. 
The multigraph $G$ is constructed starting from the ``Wheatstone bridge'' hierarchical 
lattice, described on Fig.~\ref{fig:hierarchical}.

We define the percolation problem as follows: each edge in the hierarchical lattice has multiplicity $0,1,2$ or $3$ with probability $p_0,p_1,p_2$ and $p_3=1-p_0-p_1-p_2$ (an edge with multiplicity $0$ is absent). Notice that if edges are either absent or have multiplicity $3$ (that is $p_1=p_2=0$), a subgraph is $(3,3)$ rigid if and only if it is connected. In other words, 
$(3,3)$ rigidity with triple edges is equivalent to $(1,1)$ rigidity, that is ordinary percolation, with simple edges. For the special values of the parameters $p_1=p_2=0$, this model thus contains ordinary percolation; since $(3,3)$ rigidity percolation is supposed to 
belong to a different universality class, it is interesting to study the whole phase diagram of this model. 

Let us start from a level $t$ hierarchical lattice, with $t$ large. The renormalization rule of Sec.~\ref{sec:renorm} applied to each unit cell transforms the level $t$ lattice to a level $t-1$ lattice, and induces a transformation of the parameters $p_i$:
\[
(p_0^{(t-1)},p_1^{(t-1)},p_2^{(t-1)},p_3^{(t-1)})=\varphi[(p_0^{(t)},p_1^{(t)},p_2^{(t)},p_3^{(t)})]
\]
From an analysis of the renormalization transformation, it is not difficult to obtain the explicit expression of $\varphi$, but it is very tedious.
The details are given in the Appendix. Assuming that we start from a large graph with $t\gg 1$, we iterate this renormalization transformation.
To understand the phase diagram of the model, one needs now to study the renormalization flow induced by $\varphi$; of particular interest are the fixed points of $\varphi$. Note that the four dimensional space of parameters
is actually easily reduced to three dimensions, since $p_0+p_1+p_2+p_3=1$. 

By inspection of the complicated expression for $\varphi$, three fixed points are easily found:\\
\emph{i)} $\bf{p^*}_{flop}=(p_0=1, p_1=0, p_2=0, p_3=0)$. This is the empty graph fixed point, corresponding to the floppy phase.\\
\emph{ii)} $\bf{p^*}_{rig}=(p_0=0, p_1=0, p_2=0, p_3=1)$. This is the full graph fixed point, corresponding to the rigid phase.\\
\emph{iii)} $\bf{p^*}_{perc}=(p_0=0.5, p_1=0, p_2=0, p_3=0.5)$. This is a critical fixed point, corresponding to ordinary percolation (because in this case, edges are either absent, or have multiplicity~$3$).\\

Using a Newton-Raphson scheme and scanning the whole parameter space, we have found another
fixed point:\\
\emph{iv)} $\bf{p^*}_{crit}=(p_0\simeq0.326, p_1\simeq0.174, p_2\simeq0.174, p_3\simeq0.326)$.
This fixed point may be found also by noting that the surface $p_1=p_2,~p_0=p_3$ is stable by $\varphi$. Using the normalization condition $p_0+p_1+p_2+p_3=1$, looking for a fixed point on this surface is then a one dimensional problem. We have not found any 
other fixed point in the domain $p_i\geq 0$, $p_0+p_1+p_2+p_3=1$.

The trivial fixed points $\bf{p^*}_{flop}$ and $\bf{p^*}_{rig}$ are stable.
The fixed point $\bf{p^*}_{perc}$ has three unstable directions. The fixed point $\bf{p^*}_{crit}$ has only one unstable direction.
The renormalization flow is then as follows:
\begin{itemize}
\item The three dimensional parameter space is divided by a critical hypersurface containing $\bf{p^*}_{perc}$ and $\bf{p^*}_{crit}$. On one side of the surface, the renormalization flows approaches the empty ("floppy", non percolating) fixed point; on the other side, it approaches the full fixed point ("rigid", percolating).
\item On the critical hypersurface, the flow is attracted by $\bf{p^*}_{crit}$, the "rigidity percolation" critical fixed point. 
\end{itemize}
We conclude that except for very special choices of parameters $p_1=p_2=0$, the large scale critical properties of this model are described by the "rigidity percolation" critical fixed point.

\section{Conclusion}
\label{sec:conclusion}

We have introduced and rigorously justified a renormalization transformation adapted to the study of $(k,l)$-rigidity, for $k\leq l<2k$, which generalizes the well known procedure for ordinary percolation, and the procedure for bar-joint rigidity used in \cite{pre09}.  
This method allows to solve exactly $(k,l)$-percolation problems on hierarchical graphs. We have provided such an example, 
which has the interesting feature of showing both "ordinary percolation" and "rigidity percolation" behavior for different values of the parameters.

Rather than solving exactly problems on hierarchical graphs, such renormalization transformations might be used to provide approximate solutions for problems on more realistic 2D or 3D lattices.  
This work would then provide an approximate analytical tool in the general study of $(k,l)$-percolation problems and their universality classes, a field which is still widely open. 

\bigskip

This work is supported  by the ANR-09-JCJC-009401 INTERLOP project.

\section{Appendix}

We give here the details of the computations yielding the renormalization function $\varphi$. 
The problem is simple: apply the rule of Sec.~\ref{sec:renormgraph} to an elementary ``Wheatstone bridge'' cell. Since there are $5$ edges, and each edge may be absent, single, double or triple, there are $4^5=1024$ different configurations of which we have to study the renormalization according to the rule of Sec.~\ref{sec:renormgraph}. The large number of configurations to enumerate is the only difficulty. Prop.~\ref{prop1} is useful to simplify these computations, as it allows to remove all redundant edges from the beginning.

In the following tables, we classify the configurations according to the multiplicity of their edges. For instance, an edge 
multiplicity $32210$ means that one edge has multiplicity $3$, two have multiplicity $2$, one has multiplicity $1$ and one is 
absent. Clearly, depending on how these edges are distributed on the "Wheatstone bridge", the number of renormalized edges may 
be different. The second column of the tables contains the number of edges of the renormalized cell, and the third column is the combinatorial factor corresponding to the number of configurations with the given edge multiplicity yielding the given number of renormalized edges. 

We have collected the results in 6 tables, according to the number of edges with multiplicity $3$.

\begin{table}[h!]
\title{{\bf $5$ edges with multiplicity $3$}}
\begin{tabular}{|c|c|c|}
\hline
edges multiplicity in $H$& nbr of renormalized edges & nbr configurations\\
\hline
33333 & 3 & 1\\
\hline
\end{tabular}

\end{table}

\begin{table}[ht]
\title{{\bf $4$ edges with multiplicity $3$}}
\begin{tabular}{|c|c|c|}
\hline
edges multiplicity in $H$& nbr of renormalized edges & nbr configurations\\
\hline
33332 & 3 & 5\\
33331 & 3 & 5\\
33330 & 3 & 5\\
\hline
\end{tabular}
\end{table}

\begin{table}[ht]
\title{{\bf $3$ edges with multiplicity $3$}}
\begin{tabular}{|c|c|c|}
\hline
edges multiplicity in $H$& nbr of renormalized edges & nbr configurations\\
\hline
33322 & 3 & 10\\
33321 & 3 & 20\\
33320 & 3 & 16\\33320 & 2 & 4\\
33311 & 3 & 8\\33311 & 2 & 2\\
33310 & 3 & 16\\33310 & 1 & 4\\
33300 & 3 & 8\\33300 & 0 & 2\\
\hline
\end{tabular}
\end{table}

\begin{table}[ht]
\title{{\bf $2$ edges with multiplicity $3$}}
\begin{tabular}{|c|c|c|}
\hline
edges multiplicity in $H$& nbr of renormalized edges & nbr configurations\\
\hline
33222 & 3 & 10\\
33221 & 3 & 30\\
33220 & 3 & 18\\33220 & 2 & 12\\
33211 & 3 & 24\\33211 & 2 & 6\\
33210 & 3 & 36\\33210 & 2 & 12\\33210 & 1 & 12\\
33200 & 3 & 6\\33200 & 2 & 18\\33200 & 0 & 6\\
33111 & 3 & 6\\33111 & 2 & 4\\
33110 & 3 & 6\\33110 & 2 & 12\\33110 & 1 & 12\\
33100 & 3 & 6\\33100 & 1 & 18\\33100 & 0 & 6\\
33000 & 3 & 2\\33000 & 0 & 8\\
\hline
\end{tabular}
\end{table}

\begin{table}[ht]
\title{{\bf $1$ edge with multiplicity $3$}}
\begin{tabular}{|c|c|c|}
\hline
edges multiplicity in $H$& nbr of renormalized edges & nbr configurations\\
\hline
32222 & 3 & 5\\
32221 & 3 & 20\\
32220 & 3 & 8\\32220 & 2 & 12\\
32211 & 3 & 24\\32211 & 2 & 6\\
32210 & 2 & 48\\32210 & 1 & 12\\
32200 & 2 & 12\\32200 & 1 & 12\\32200 & 0 & 6\\
32111 & 2 & 20\\
32110 & 2 & 12\\32110 & 1 & 48\\
32100 & 2 & 12\\32100 & 1 & 12\\32100 & 0 & 36\\
32000 & 2 & 4\\32000 & 0 & 16\\
31111 & 1 & 5\\
31110 & 1 & 12\\31110 & 0 & 8\\
31100 & 1 & 12\\31100 & 0 & 18\\
31000 & 1 & 4\\31000 & 0 & 16\\
30000 & 0 & 5\\
\hline
\end{tabular}
\end{table}

\begin{table}[ht]
\title{{\bf No edge with multiplicity $3$}}
\begin{tabular}{|c|c|c|}
\hline
edges multiplicity in $H$& nbr of renormalized edges & nbr configurations\\
\hline
22222 & 3 & 1\\
22221 & 3 & 5\\
22220 & 2 & 5\\
22211 & 2 & 10\\
22210 & 1 & 20\\
22200 & 1 & 6\\22200 & 0 & 4\\
22111 & 1 & 10\\
22110 & 1 & 6\\22110 & 0 & 24\\
22100 & 1 & 6\\22100 & 0 & 24\\
22000 & 1 & 2\\22000 & 0 & 8\\
21111 & 0 & 5\\
21110 & 0 & 20\\
21100 & 0 & 30\\
21000 & 0 & 20\\
20000 & 0 & 5\\
11111 & 0 & 1\\
11110 & 0 & 5\\
11100 & 0 & 10\\
11000 & 0 & 10\\
10000 & 0 & 5\\
00000 & 0 & 1\\
\hline
\end{tabular}
\end{table}

\newpage

Collecting the information from these tables, we obtain the expression for the renormalization function $\varphi$. Calling the renormalized probabilities $p'_0,p'_1,p'_2,p'_3$, we have
\begin{eqnarray}
p'_3&=& =p_3^5+5p_3^4p_0+8p_3^3p_0^2+2p_3^2p_0^3+p_2^5+5p_2^4p_1+6p_3^2p_1p_0^2+6p_3^2p_1^2p_0+4p_3^2p_1^3 \nonumber \\
&&+16p_3^3p_1p_0+8p_3^3p_1^2+5p_3^4p_1+5p_3p_2^4+8p_3p_2^3p_0+6p_3^2p_2p_0^2+18p_3^2p_2^2p_0
\nonumber\\
&&+10p_3^2p_2^3+10p_3^3p_2^2+16p_3^3p_2p_0+5p_3^4p_2+20p_3^3p_2p_1+30p_3^2p_2^2p_1+24p_3^2p_2p_1^2\nonumber \\
&&+36p_3^2p_2p_1p_0+20p_3p_2^3p_1+24p_3p_2^2p_1^2 \nonumber \\
p'_2 &=& 10p_2^3p_1^2+12p_3^2p_1^2p_0+6p_3^2p_1^3+2p_3^3p_1^2+4p_3p_2p_0^3+12p_3p_2^2p_0^2+12p_3p_2^3p_0\nonumber \\
&&+18p_3^2p_2p_0^2+4p_3^3p_2p_0+6p_3^2p_2p_1^2+12p_3^2p_2p_1p_0+6p_3p_2^2p_1^2+48p_3p_2^2p_1p_0\nonumber\\
&&+20p_3p_2p_1^3+12p_3p_2p_1^2p_0+12p_3p_2p_1p_0^2+5p_2^4p_0+12p_3^2p_2^2p_0\nonumber \\
p'_1 &=& 2p_2^2p_0^3+6p_2^3p_0^2+10p_2^2p_1^3+6p_2^2p_1^2p_0+6p_2^2p_1p_0^2+20p_2^3p_1p_0+4p_3p_1p_0^3+12p_3p_1^2p_0^2\nonumber \\
&&+12p_3p_1^3p_0+5p_3p_1^4+18p_3^2p_1p_0^2+12p_3^2p_1^2p_0+4p_3^3p_1p_0+12p_3p_2^2p_0^2+12p_3^2p_2p_1p_0\nonumber \\
&&+12p_3p_2^2p_1p_0+48p_3p_2p_1^2p_0+12p_3p_2p_1p_0^2\nonumber \\
p'_0 &=& p_0^5+5p_3p_0^4+8p_3^2p_0^3+2p_3^3p_0^2+5p_2p_0^4+8p_2^2p_0^3+4p_2^3p_0^2+p_1^5+5p_1^4p_0+10p_1^3p_0^2\nonumber\\
&&+10p_1^2p_0^3+5p_1p_0^4+5p_2p_1^4+20p_2p_1^3p_0+30p_2p_1^2p_0^2+20p_2p_1p_0^3+24p_2^2p_1^2p_0\nonumber\\
&&+24p_2^2p_1p_0^2+16p_3p_1p_0^3+18p_3p_1^2p_0^2+8p_3p_1^3p_0+6p_3^2p_1p_0^2+36p_3p_2p_1p_0^2
\nonumber \\
&&+6p_3p_2^2p_0^2+6p_3^2p_2p_0^2+16p_3p_2p_0^3 \nonumber
\end{eqnarray}

\end{document}